\UseRawInputEncoding
\documentclass{optica-article}

\journal{opticajournal} % for journals or Optica Open

\articletype{Research Article}

\usepackage{lineno}

\begin{document}

\title{The Problem of No Return Photon Ranging Measurements with Entangled Photons}

\author{Mohit Khurana\authormark{1,2,*}}
\address{\authormark{1}Institute for Quantum Science and Engineering, Texas A\&M University, College Station, TX 77843, USA\\
\authormark{2}Department of Physics and Astronomy, Texas A\&M University, College Station, TX  77843, USA\\}
\email{\authormark{*}mohitkhurana@tamu.edu}

\begin{abstract*}We introduce a fascinating problem of light detection and ranging measurement without necessitating the return of the photon directed towards the target or object. We approach this challenging problem using quantum entanglement - an entangled pair of photons; one photon is sent toward the target or object, while the other is directed into a medium, which undergoes continuous measurements. We assume the light-matter interaction at the target such that the quantum state collapse is probabilistically biased. We present thought experiments and measurement schemes to conduct correlation measurements and examine the methodology of these measurements to estimate the target's range, using a maximally entangled Bell state ($|\Psi^+\rangle = \frac{1}{\sqrt{2}} (|H_1V_2\rangle + |V_1H_2\rangle)$) as an example. When the photon interacts with the target, the Bell state undergoes biased decoherence or state collapse, leaving a signature in the spatial correlation G(x) quantity.
\end{abstract*}

\section{Introduction}
\noindent Over the last hundred years, a wide range of scientific tools based on principles of quantum mechanics have transformed fields like chemistry, biosciences, metrology, sensing, imaging, and computing by providing capabilities that exceed classical boundaries. For instance, in metrology, quantum principles like superposition, entanglement, and squeezed states of light enhance precision, allowing atomic clocks and interferometers to achieve unprecedented accuracy for applications such as sensing, navigation, and gravitational wave detection \cite{Giovannetti2004, Aslam2023, Aasi2013}. Quantum teleportation, quantum communications, and quantum networks are key components of quantum information science, which leverages principles of quantum mechanics (like superposition and entanglement) to transmit, process, and secure information in ways that classical systems cannot.\cite{Bouwmeester1997,Salih2013, Duan2001, Kimble2008}. Quantum computing, which uses entangled states, can perform complex calculations exponentially faster than classical computers to solve problems in cryptography, optimization, and materials sciences \cite{Ladd2010}. These advancements continue to push technological boundaries, unlocking new frontiers in science and engineering. In this work, we introduce a novel problem and explore another significant application based on the core principles of quantum mechanics -- quantum entanglement between two pairs of photons to measure the distance between the source and target without requiring the photon return from the target. 
\begin{figure}[ht!]
\centering\includegraphics[width=7cm]{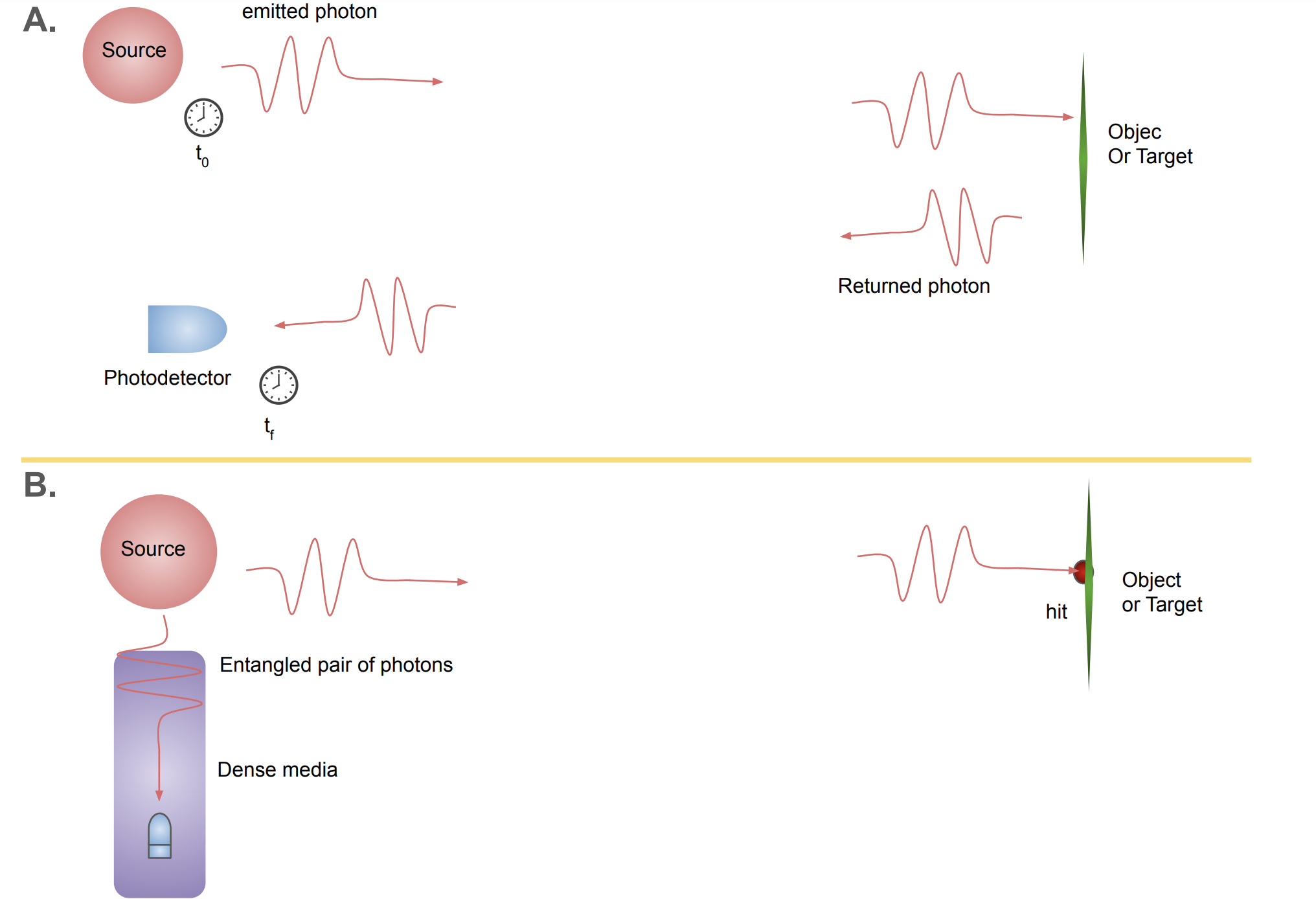}
\caption{A). Range measurement scheme with the return of photons from the target to the detector. B). Thought experiment measurement scheme of no return photon range measurement with an entangled pair of photons. In the measurement setup for a second photon, the photon travels through a medium (e.g., vacuum, dense medium, fiber, or waveguide), and a signal is detected when the first photon interacts with the target.}
\label{A_and_B}
\end{figure}
Ranging and imaging measurement techniques using photons, such as light detection and ranging systems, rely on the returned photons and efficient collection of returned photons to accurately measure distances and generate detailed 3D representations of environments. Two critical aspects of these systems are photon collection efficiency (CE) and external light source noise in the signal. CE ($\eta$) refers to the proportion of emitted photons that are successfully detected after interacting with a target i.e., $\eta = \frac{\text{N}_{\text{detected}}}{\text{N}_{\text{emitted}}}$. Factors that affect $\eta$ include the receiver aperture size, alignment, target reflectivity, light scattering, and the distance between the detector and the target or object. The measurement involves efficient detection schemes (such as direct or heterodyne), noise removal, and signal filtering. The inverse square law - light intensity diminishing with the square of the distance from the source limits the distance of ranging measurements. These challenges motivate us to think about ranging measurements with light without the return of photons or the collection of returned photons from the target or object to the detector. A schematic diagram is shown in Fig. \ref{A_and_B} to compare return and no return photon ranging measurements with light.
\\

\section{Measurement schemes and correlation}
\noindent When two photons are entangled, measuring one of them affects the entire system due to the collapse of the wavefunction. For example, if two photons are entangled in polarization, measuring the polarization of one photon will instantaneously determine the polarization of the other. Standard quantum measurement, described by the projection postulate, is inherently destructive to superposition and entanglement. First, we propose a thought experiment involving a pair of quantum-entangled photons to exchange information between two photons when the first photon (right-moving) interacts with the target. A pair of entangled photons is created from a source; one photon is directed at the target or object, while the other photon goes to the measurement setup. The measurement of the second photon (down-moving) reveals the instant the first photon hits the target.
\\

\begin{figure}[ht!]
\centering\includegraphics[width=15.5cm]{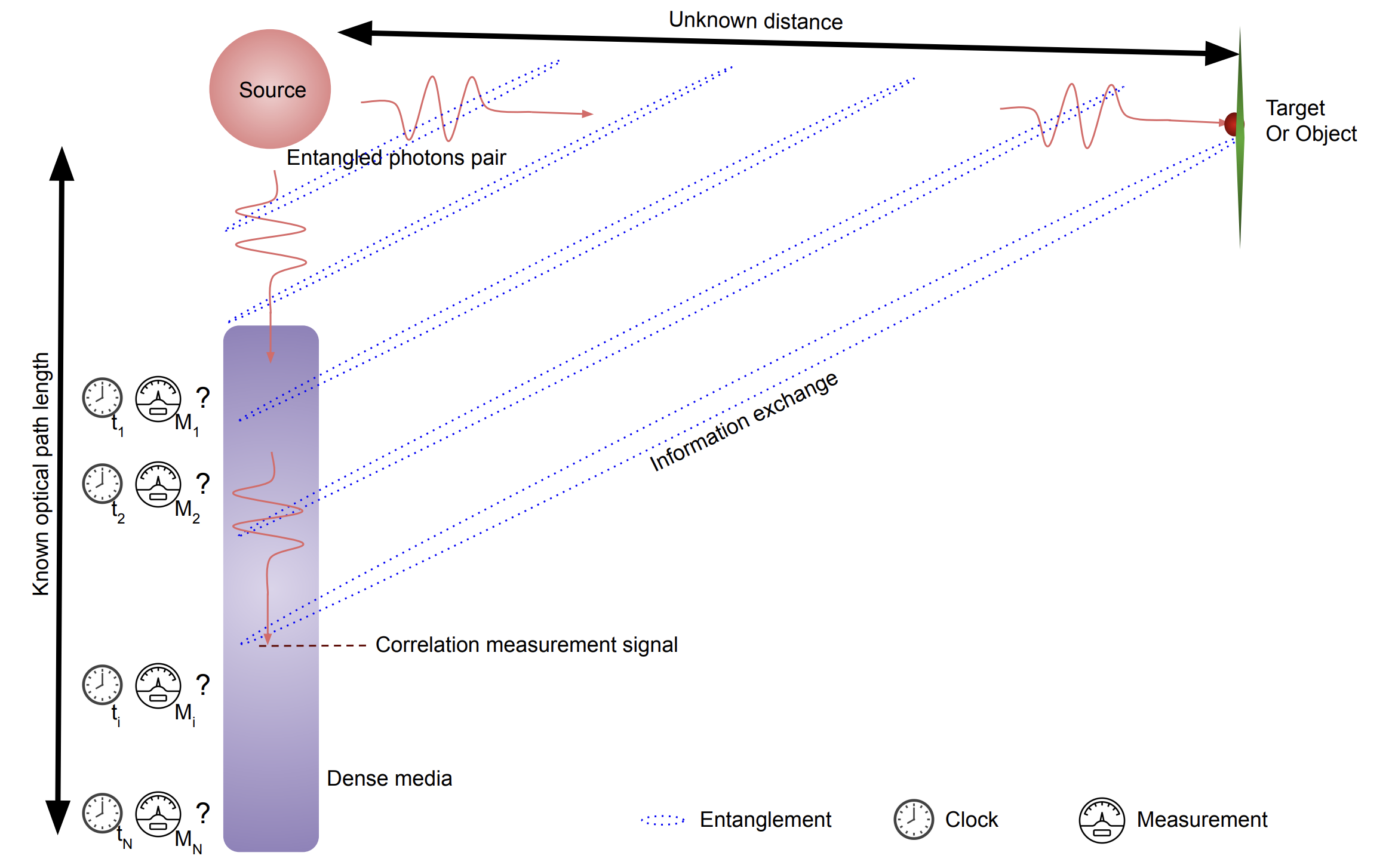}
\caption{A source creates a pair of photons in a maximally entangled Bell state. The first photon (right-moving) is sent towards the target, and the second photon (down-moving) travels through a medium (vacuum or dense media). While the second photon travels, continuous measurements are made. When the first photon hits the target (or object), due to the collapse of the state and exchange of information with the second photon, the correlation measurement signal reveals a signature at that instant. The time delay or difference and optical path can be calculated with a clock at all measurements for the second photon.}
\label{entanglement_1}
\end{figure}

\noindent It is essential to note that in a typical entangled photon system, detecting entanglement requires correlation measurements between both photons. However, if the first photon is not accessible and only measurement access to the second photon is possible, direct correlation measurements involving both photons are not possible. In typical Bell states experiments, correlation tests are conducted that involve both photons. However, in our case, since access to the first photon is not possible, traditional Bell-type correlation tests cannot be performed. Instead, we can analyze statistical correlations in the measurement outcomes of the second photon over multiple trials to determine whether the entanglement properties are still preserved. This approach relies on quantum coherence effects, which disappear if a measurement has already collapsed the entangled state. We define operator $O_2$ observable for a measurement on the second photon, and consider continuous measurements of second photon state as it travels through the measurement setup shown in Fig. \ref{entanglement_1}. The statistical correlations, such as higher-order moments of measurement outcomes, fluctuations in variance of measurement results, or temporal correlations between measurements of the second photon, can indicate whether quantum coherence is maintained between the two photons and if the first photon interacts with the target, leading to a modification or collapse of the state (see Fig. \ref{entanglement_1}). 
\\

\begin{figure}[ht!]
\centering\includegraphics[width=15.5cm]{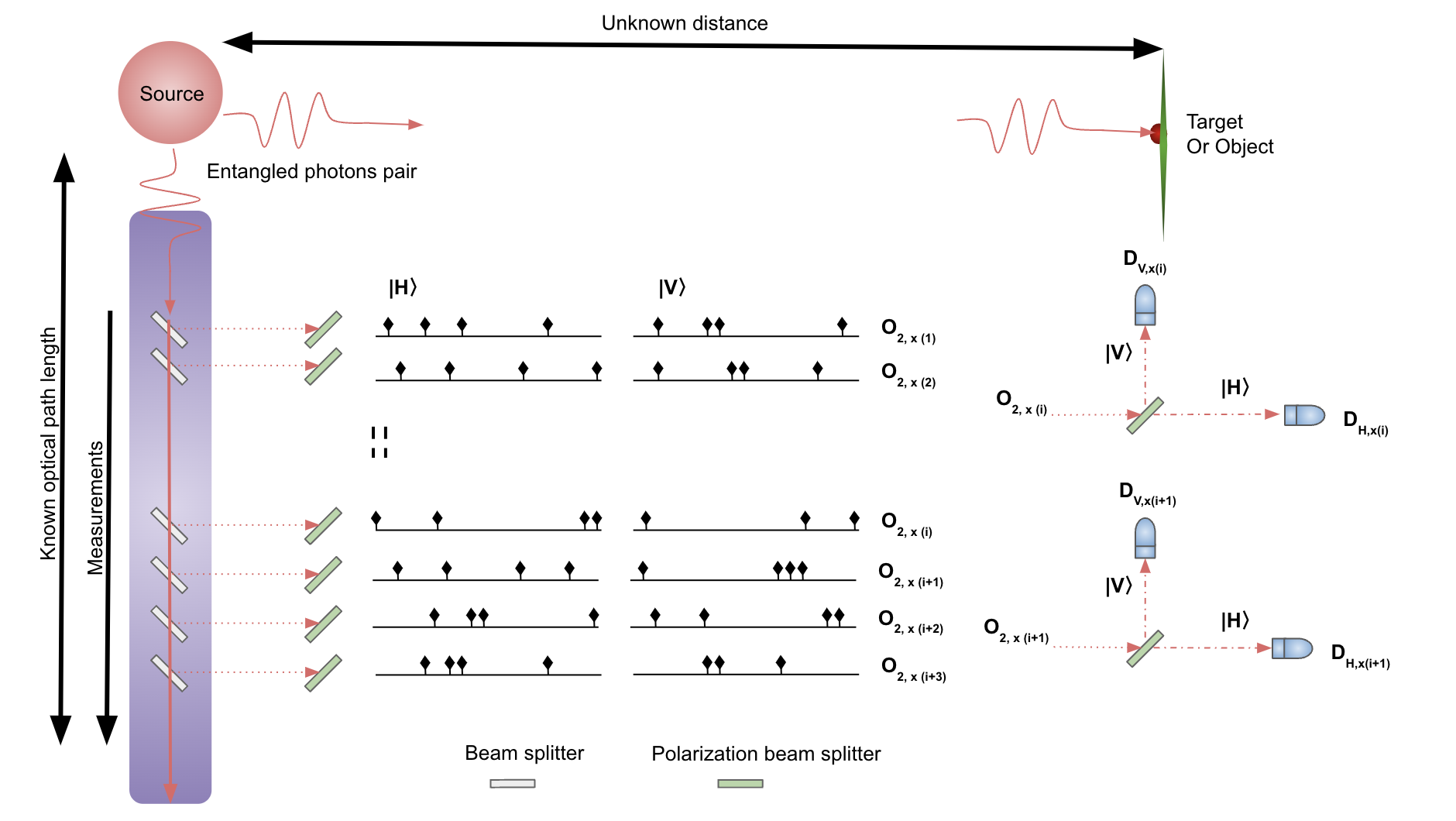}
\caption{Measurement scheme I: A source generates a maximally entangled Bell state; one photon is sent toward the target, and the other travels through the measurement setup. Measurements ($O_{2}$) at locations ($x_i$) are performed with a beam splitter and a PBS. The photons are annihilated at the detectors to count the number of photons in each polarization state.}
\label{entanglement_2}
\end{figure}

\noindent Higher-order moment correlation function for an observable \( O_2 \) on the second photon, $C_n = \langle O_2^n \rangle - \langle O_2 \rangle^n$ would exhibit non-classical correlations if entanglement is preserved. A deviation of \( C_n \) from classical predictions indicates that quantum coherence (and thus entanglement) remains intact. Even though we have access to the second photon, we can check whether its measurement variance behaves dynamically as if it is still entangled. The variance of the measurement results is given by, $\sigma^2(O_2) = \langle O_2^2 \rangle - \langle O_2 \rangle^2$.  Tracking \( \sigma^2(O_2) \) over multiple trials, if entanglement is preserved, then the variance remains large due to quantum interference effects. In temporal correlation, $G(\tau) = \langle O_2(t) O_2(t+\tau) \rangle - \langle O_2(t) \rangle \langle O_2(t+\tau) \rangle$, exhibits long-range quantum correlations if entanglement is preserved; otherwise, it shows short-range or classical correlations. In our situation, a time correlation quantity can be measured in general, but we must simplify the time dependency complexity resulting from the random time intervals between two entangled photon pairs from the source. Additionally, measuring time-dependent correlation requires more photons and stricter conditions on the photon detection to obtain adequate statistics. Therefore, we must perform a correlation measurement that is space ($x_i$) dependent with no time dependence. So, we introduce correlation measurement G$(x_i)$, where $x_i$ is either continuous or discrete variable (see Fig. \ref{entanglement_2}),
\begin{equation}
    G_{x}=\frac{\langle O_{2, x+dx} O_{2,x}\rangle}{\langle O_{2, x+dx}\rangle \langle O_{2,x}\rangle} ; G_{x(i)}=\frac{\langle O_{2, x(i+1)} O_{2,x(i)}\rangle}{\langle O_{2, x(i+1)}\rangle \langle O_{2,x(i)}\rangle}
\end{equation}

\noindent It is important to note that we have considered the interaction of light with the target as a general case. The target may include materials such as wood, metal, glass, paint, concrete, liquids, or gases, among others. Moreover, the degree of freedom must be carefully selected for each specific type of target; this can be explored in detail through experiments with various types of targets, which is beyond the scope of this work. Since the interaction of light with the target is unknown, the measurement scheme needs to be appropriately adopted with other degrees of freedom related to the entangled properties of photons to achieve accurate results. For the purposes of this discussion, we limit our focus to one degree of freedom: polarization. Therefore, we evaluate our outcome in probabilistic terms regarding the polarization state of the photon as a degree of freedom in this context. We consider an initial maximally entangled Bell state in terms of polarization,  $|\psi\rangle = \frac{1}{\sqrt{2}} (|H\rangle_1 |V\rangle_2 + |V\rangle_1 |H\rangle_2)$, where $|H\rangle$ and $|V\rangle$ denote horizontal and vertical polarization states, respectively. The subscripts 1 and 2 refer to the first (right-moving) and second (down-moving) entangled photons (see Fig. \ref{entanglement_2}). When the first photon hit the target or interacts with it, the quantum state of the system undergoes a transformation into a probabilistic mixture of possible outcomes, represented by the density matrix, $\rho = A | \psi_{\text{lost}} \rangle \langle \psi_{\text{lost}} | + B | \psi_{\text{mixed}} \rangle \langle \psi_{\text{mixed}} | + C | \psi_{\text{coherent}} \rangle \langle \psi_{\text{coherent}} |$, where \( A \), \( B \), and \( C \) are the probabilities (with \( A + B + C = 1 \)) associated with each outcome. Here, \( | \psi_{\text{lost}} \rangle \) represents the state where the photon has been absorbed or lost to the system, with no further information about the photon's state. \( | \psi_{\text{mixed}} \rangle \) describes a mixed state, where the photon is in a probabilistic combination of different outcomes due to factors like decoherence or partial measurement. \( | \psi_{\text{coherent}} \rangle \) represents a coherent state, in which the photon retains a well-defined superposition, maintaining phase coherence. In case of absorption, the photon is lost, leaving the second photon in a completely mixed state,  $|\psi_{\text{lost}}\rangle = I \otimes \rho_{\text{mix}}$ \cite{Nielsen2012}, in case of depolarization or scattering, the photon is depolarized but still present, $|\psi_{\text{mixed}}\rangle = \frac{1}{\sqrt{2}}(|H\rangle_1 |H\rangle_2 + |V\rangle_1 |V\rangle_2)$ and in case of specular reflection, the partial entanglement is retained with a phase shift, $|\psi_{\text{coherent}}\rangle = \frac{1}{\sqrt{2}}(|H\rangle_1 |V\rangle_2 + e^{i\phi} |V\rangle_1 |H\rangle_2)$ \cite{Nielsen2012}.
\\

\noindent Here, we use the $O_2$ measurement where only a tiny probability of measurement exists using an asymmetric beam splitter and reflected photons polarization state is detected with a polarization beam splitter and detectors, i.e., if N entangled photons are traveling through the measurement setup, only a small proportion of total N entangled photons are detected and rest of the photons remains untouched. If the first photon remains unmeasured, the second photon should show correlations consistent with entanglement. When a measurement is conducted on the first photon, we should note a sudden change in the measurement outcomes of the second photon, signaling the change in entanglement properties. By fixing the distance or optical path between the source and the measurement locations of the second photon and recording the timing of measurements, we can assess the total time the first photon takes to reach the target and thus evaluate the range when the correlations show significant change. 
\\

\noindent The system starts in the maximally entangled Bell state, $|\psi\rangle = \frac{1}{\sqrt{2}} (|H\rangle_1 |V\rangle_2 + |V\rangle_1 |H\rangle_2)$. This state exhibits perfect quantum anti-correlation (opposite outcomes) in the \(H/V\) basis and maintains entanglement in all polarization bases. We consider sending N photon pairs one by one, with random intervals between each pair, but a large number of pairs within a short period relative to the total time taken by the first photon to reach the target, thus ensuring fast measurement. In the measurement scheme shown in Fig. \ref{entanglement_2}, we count the number of photons that are either in H or V polarization state, so we define $O_{2}$ such that an asymmetric beam splitter with a small reflection coefficient (for instance, R $\sim$ 0.001) and a polarization beam splitter (PBS) that determines the polarization of the second photon and the correlation measurement quantity as,

\begin{equation}
    G_{H,x(i)} = \frac{\langle \hat{a}_{H,x(i)}^\dagger \hat{a}_{H,x(i)} \rangle \cdot \langle \hat{a}_{H,x(i+1)}^\dagger \hat{a}_{H,x(i+1)} \rangle}{\langle \hat{a}_{H,x(i)}^\dagger \hat{a}_{H,x(i)} \rangle + \langle \hat{a}^\dagger_{H,x(i+1)} \hat{a}_{H,x(i+1)} \rangle}
\end{equation}
\noindent where $\hat{a}_{H,x(i)}^\dagger$, $\hat{a}_{H,x(i)}$, $\hat{a}_{H,x(i+1)}^\dagger$ and $\hat{a}_{H,x(i+1)}$ are creation and annihilation operator for H polarization state at $x$ and $x(i+1)$ locations, respectively. We define $G_{x(i)}$ to estimate the deviation in either of H or V polarization state of photon,

\begin{equation}
    G_{x(i)} = |G_{H,x(i)} - G_{V,x(i)}|
\end{equation}

\noindent therefore,
\begin{equation}
    G(O_{2, x(i)}) = G_{x(i)} = \Bigg| \frac{n_{H,O_{2, x(i)}} * n_{H,O_{2, x (i+1)}}}{n_{H,O_{2, x(i)}} + n_{H,O_{2, x(i+1)}}}  - \frac{n_{V,O_{2, x(i)}} * n_{V,O_{2, x(i+1)}}}{n_{V,O_{2, x(i)}} + n_{V,O_{2, x(i+1)}}}\Bigg|
\end{equation}

\noindent where $n_{H,O_{2, x(i)}}$, $n_{V,O_{2, x(i)}}$, $n_{H,O_{2, x(i+1)}}$, and $n_{V,O_{2, x(i+1)}}$ represent the number of photons measured in H and V polarization at locations $x(i)$ and $x(i+1)$, respectively. This correlation function $G(x(i))$ quantifies the polarization consistency of photons between two adjacent measurement locations ($x(i)$). It measures the deviation between the normalized products of horizontally and vertically polarized photon counts at consecutive positions. A higher  $G(x(i))$  value indicates a difference in polarization retention across adjacent measurements, suggesting that the polarization state fluctuates significantly. Deviations in $G(x(i))$ indicate interference, decoherence, or environmental effects influencing polarization evolution. When the first photon hits the target or object, we seek to evaluate $G_{hit} = max(G(O_{2, x(i)})); \forall i$. The signal to noise ratio for $G(x(i))$ or $G_{hit}$ $\sim$ total number of photons (or $N_{emitted}/2$). 
\\

\noindent When the first photon undergoes depolarization due to interaction with the target medium, its polarization state becomes randomized, leading to a mixed-state description rather than a pure state. This transformation alters the system into a statistical mixture, $\rho_{\text{mixed}} = \frac{1}{2} (|H_1H_2\rangle \langle H_1H_2| + |V_1V_2\rangle \langle V_1V_2|)$. i.e. the system has a  equal 50\% - 50\% probability of being in \(|H_1H_2\rangle\) and \(|V_1V_2\rangle\), respectively but no longer in a quantum superposition. The depolarization eliminates quantum entanglement and reduces the system to classical statistical behavior where photons behave independently in all measurement bases. In correlation quantity G, we can only observe a signal of a hit when there is a biased polarization state outcome of second photons. Without loss of generality, we simplify this problem to two variables, a and b, where a and b represent the probabilities of the second photon being in the H and V states, respectively, after the first photon interacts with the target. For the correlation measurement G defined above, in the case of a = b, there is no way to determine if the first photon hit the target. However, in the case of a $\neq$ b, the G quantity would provide a signal to measure the range of the object or target when the first photon makes contact. As long as $a$ and $b$ are not equal, this provides a sufficient condition for the measurement to determine if the target or object has been hit, although a high number of photons are required to increase the signal-to-noise ratio of $G_{hit}$. 
\\

\begin{figure}[ht!]
\centering\includegraphics[width=15.5cm]{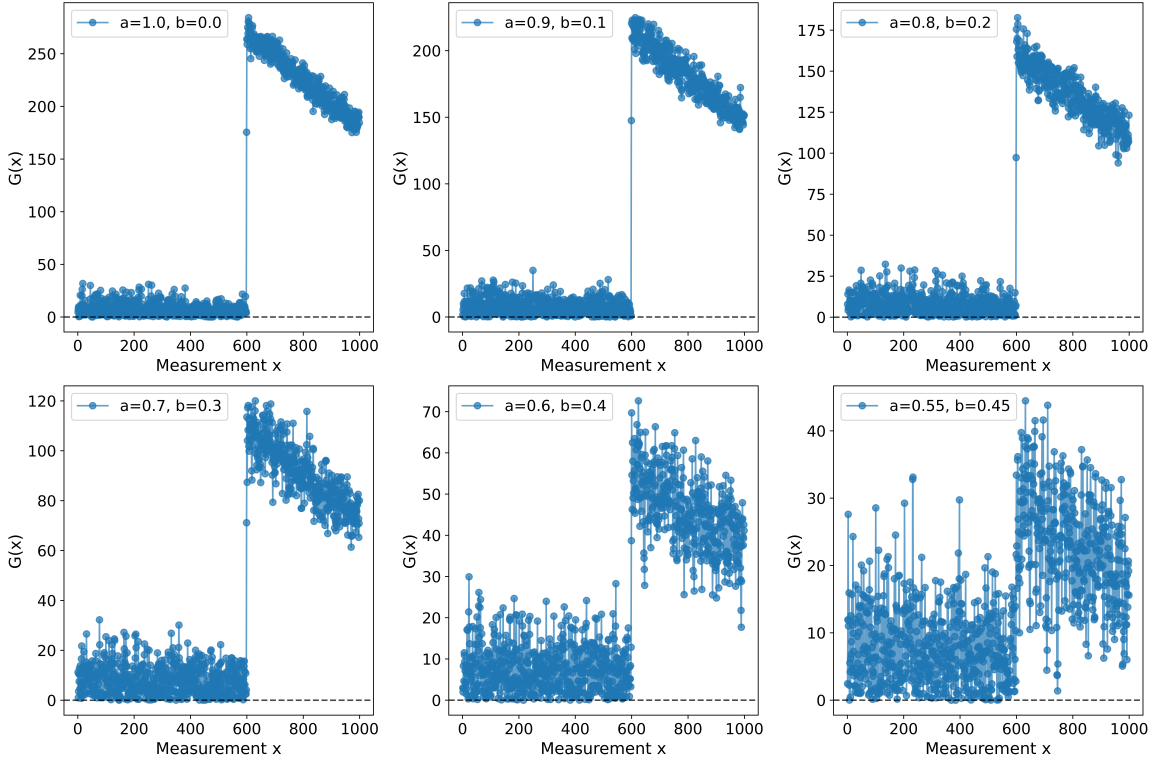}
\caption{Numerical experiment results of our first measurement scheme -  using a maximally entangled Bell state and measurements at $x_i$ locations on the second photon. We assume that as the first photon hits that target or object, the state or wavefunction collapses into the $|H\rangle$ (or $|V\rangle$) state. $10^6$ photons (one by one) sent towards the down-moving detection process or scheme, at a location $x_i$ = 600, the photons are in either H or V polarization due to the first photon hit at the target (the state collapse in H (or V) polarization, i.e., all the photons that hit collapses into same polarization). Swapping a and b would yield similar results due to G's symmetry for polarization H or V state.}
\label{coorelation_subplots}
\end{figure}

\noindent  Now, we conduct a numerical experiment to demonstrate how we can find the information of the first photon hit from the correlation measurements. Suppose we start with 2*$10^6$ entangled photon pairs and a total of $10^3$ measurements (each measurement includes a beam splitter with reflection $\sim$ 0.001 and a PBS). The optical path length for the second photon at $x(i)$ = 600 equals when the first photon hits the target and assumes that the interaction of the photon at the target is such that the polarization state of the first photon collapses to  H (or V) (i.e., all the photons collapse into H (or V) polarization). Thus, we can impose a constraint - forcing all second photons to be V (or H) polarized after \( x \geq 600 \). We consider random polarization state of second photon before $x(i)=600$,  see Fig. \ref{coorelation_subplots}, with different values of a and b. Fig. \ref{coorelation_subplots} shows the drastic increase of $G(x(i))$, highlighting that polarization variation has been eliminated beyond this point, and we have find when the first photon hit the target. Note that we have assumed a + b = 1 in our numerical simulation; however, in general, we can have a + b $\leq$ 1. Nevertheless, more photons would be needed for small values of a and b to achieve a significant signal-to-noise ratio in correlation measurement and estimate the target range. Note that if there is a significant penetration depth across the biasing state collapse ($a\neq b$), we would observe multiple peaks in G(x).
\\
% \clearpage

\begin{figure}[ht!]
\centering\includegraphics[width=15.5cm]{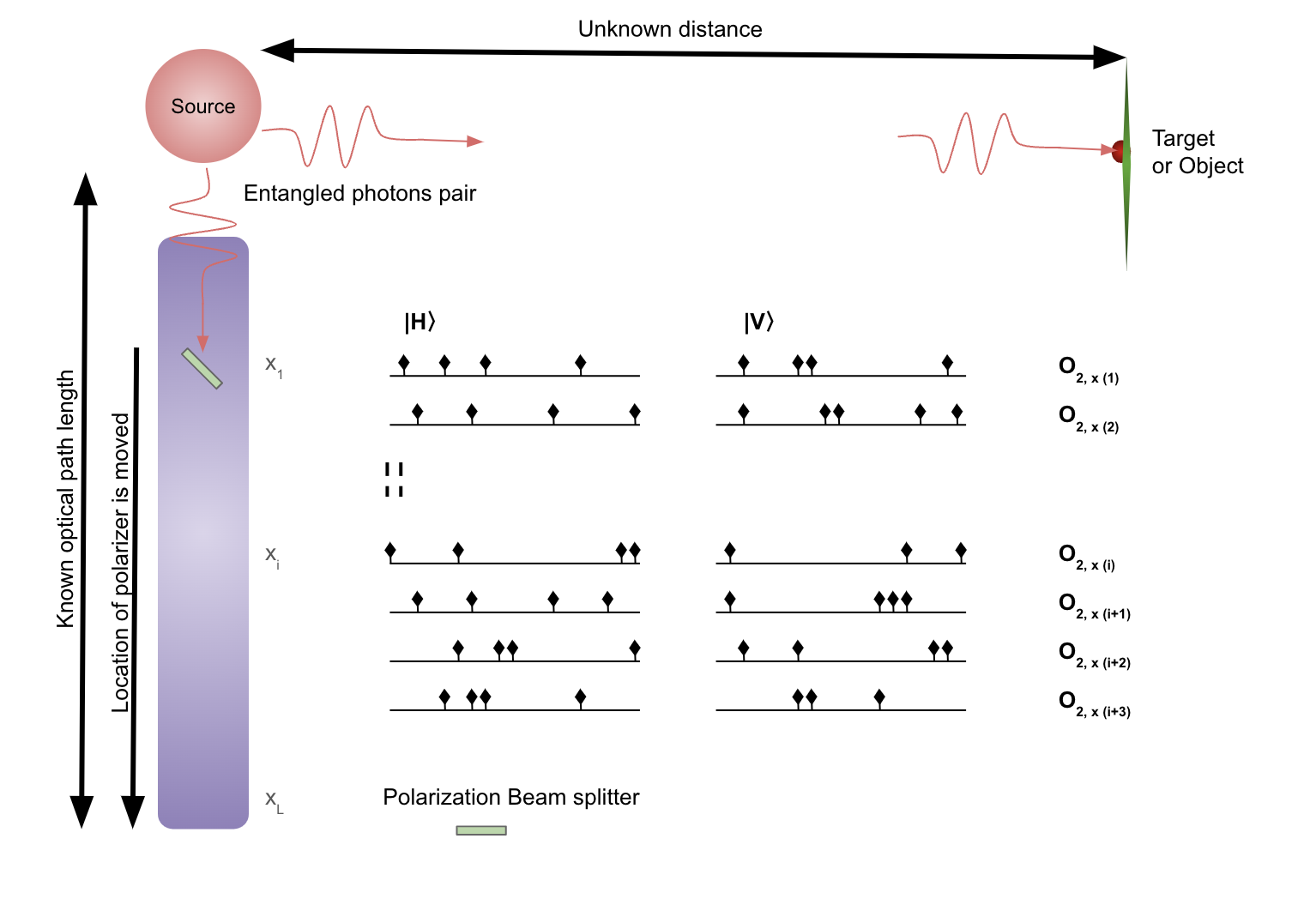}
\caption{Measurement scheme II: Moving PBS in dense media to record the number of photons in H or V polarization state. A PBS is moved along the system, and measurements are performed at locations x$_i$.}
\label{moving_polarizer}
\end{figure}

\noindent The first measurement scheme of counting the photons in each polarization can be modified appropriately to reduce the total number of photons needed to measure the correlation G($x(i)$). By moving a PBS along with measurement locations $x(i)$ over the same time interval for a single measurement at a location in a periodic fashion and adjusting their positions to register at all locations, subsequently, we measure the number of photons in each polarization at all locations to calculate G$(x(i))$. We assume that the time interval between two pairs is much smaller than a pulse, and the pulse width is also much smaller than the total time travel of the first photon hitting the target. We have shown this scheme in Fig. \ref{moving_polarizer}, where only one PBS is used, and its location with the photon detectors would collect the signal required to measure G$(x(i))$.
\\

\noindent Slow-light structures such as photonic crystal waveguides (PhCW) can support both H and V polarization with very slow group velocity. PhC integrated with an overlaid dielectric waveguide to extract a small portion of the propagating light via evanescent coupling, ensuring controlled leakage without significant loss for both polarizations. The overlaid waveguide, carefully designed for phase matching and optimal separation, guides the extracted light to an on-chip PBS. The on-chip PBS separates H and V polarization into distinct waveguide paths. The split polarization components are then directed to independent on-chip waveguide integrated and coupled avalanche photodiodes or single-photon detectors to detect the number of photons via electronics. This fully integrated system fabricated using silicon photonics or III-V materials, can enable real-time polarization-resolved photon detections for our measurement schemes.
\\

\noindent To create a tunable coupling system that mimics a moving on-chip PBS, a series of optical couplers can be utilized, where the coupling strength of each coupler is dynamically adjusted. This can be achieved by incorporating tunable elements, such as electro-optic materials. The coupling strengths are controlled to vary sequentially, starting with the first coupler at high coupling strength and gradually reducing it to zero. Meanwhile, the second coupler starts with low coupling strength and increases to high, while the third coupler follows a similar pattern, i.e. changing coupling coefficients over time, with each coupler transitioning from high to low coupling strength in a periodic fashion. This shifting sequence of coupling strengths creates the effect of a moving on-chip PBS, where the polarizing action appears to move along the system. Nevertheless, various measurement setup configurations with integrated photonics and optical components can be arranged on an optical bench to conduct the experiment.
\\

\noindent Suppose we need to measure a range of up to 100 m, with a resolution of 0.1 - 0.01 m in depth. To obtain this, we would need $10^{3}$ - $10^{4}$ measurements (x$_{i}$) for the second photon. We need approximately $10^2$ - $10^4$ photons per measurement for accurate statistics, resulting in a requirement of about 10$^6$ - 10$^8$ photons overall. This photon requirement is several orders of magnitude lower than the known methods to date; the latest LiDAR operates at nearly 10$^{13}$ photons per measurement. Note that if the target range is bounded, the number of measurements can be reduced, which would lower the number of photons required in the scheme; for instance, target is within the range of 90-100 m, we can bound the min and max locations of measurements ($x_k$ - $x_l$, where it corresponds to equal optical path length of 90-100 m distance travel for first photon), thus requiring a low number of photons for correlation measurements.
\\

\begin{figure}[ht!]
\centering\includegraphics[width=15.5cm]{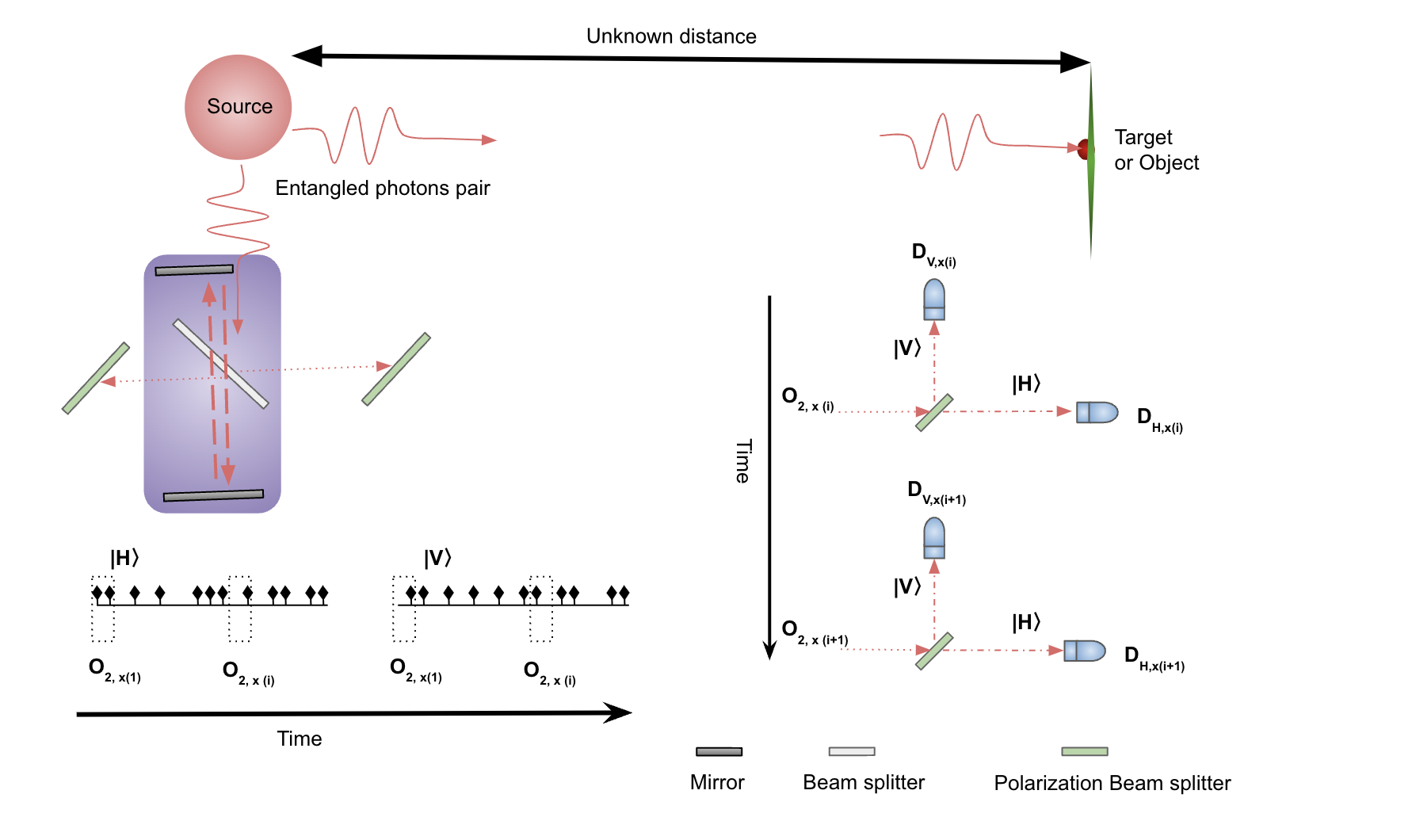}
\caption{Measurement scheme III: A setup with a mirror at the bottom and top and an asymmetric beam splitter (R $\sim$ 0.001), which are aligned at some angle, with incoming photon direction. The reflected photons are sent toward a PBS to detect the number of photons in the H or V polarization state. The pairs are sent in a pulse where the pulse time width is much shorter than the time travel by photons between the beam splitter and mirror. Each measurement $O_{2,x(i)}$ is triggered at the time to record all the photons collected in detectors for each beam reflection at the beam splitter. This measurement setup is compared with another setup shown in Fig \ref{two_setups}.}
\label{compact_measurement}
\end{figure}

\noindent Now, we discuss our third measurement scheme, which is shown in Fig. \ref{compact_measurement}. In this scheme, the distance between the center asymmetric beam splitter and the two ends, namely the top and bottom mirrors, is adjusted to appropriately record the number of photons in each measurement during a time interval, completing the ith measurement, which corresponds to the x$_i$ measurement of the scheme I and II. Proper time triggering and measurement recording of many photons in a small interval gives us one measurement for that interval; this interval is small compared to the time a photon travels between the middle beam splitter and the bottom mirror. This third scheme is equivalent to I and II measurement schemes but is compact and requires only a few optical components, with each successive measurement done using the same optical elements.
\\

\begin{figure}[ht!]
\centering\includegraphics[width=10cm]{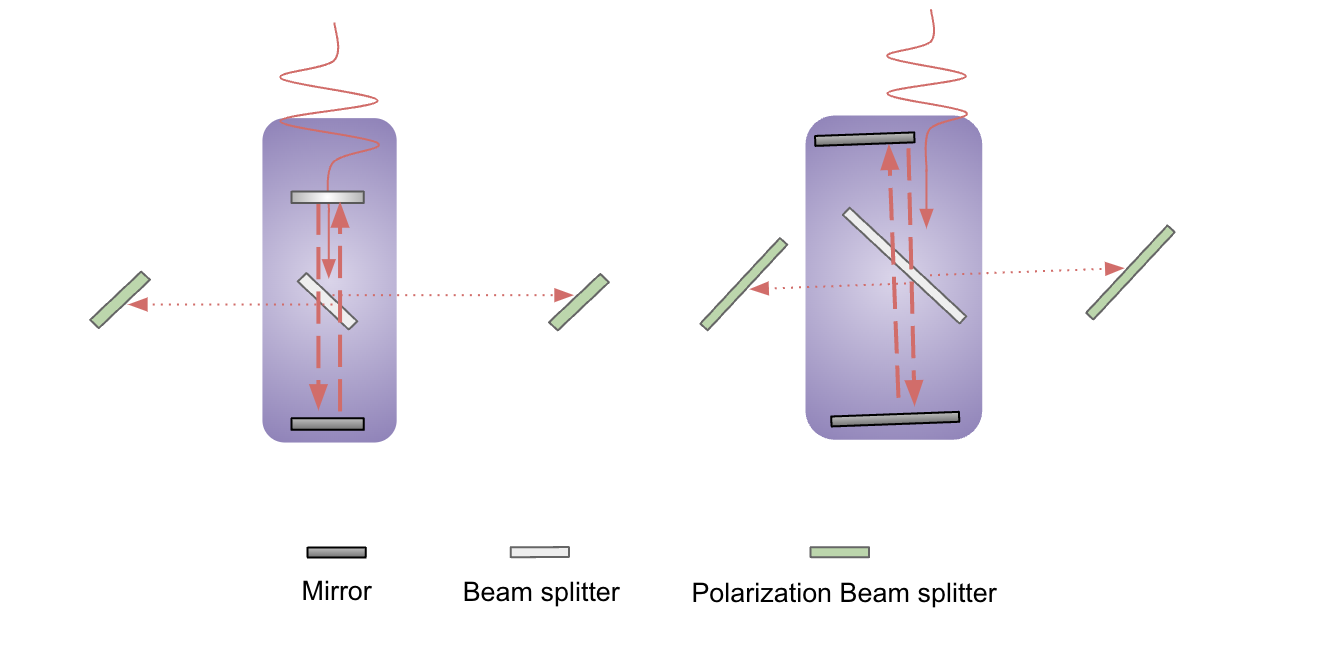}
\caption{The left configuration features a standard beam splitter on top. In contrast, the right configuration includes a mirror on top, with the system aligned at an angle. The right configuration design facilitates the entry of photons and enables reflections between the top and bottom mirrors, maximizing the time the photons stay within the system.}
\label{two_setups}
\end{figure}

\noindent Fig. \ref{two_setups} shows two different measurement setups described in scheme III. In the left setup, a beam enters from the top and passes through a standard beam splitter, partially transmits and reflects the light, followed by an asymmetric beam splitter reflecting a small portion of light, R$\sim$0.001. The transmitted portion of photons travels downward, gets reflected by a mirror at the bottom, and continues this back-and-forth motion. In contrast, the setup on the right is slightly tilted, and we assume the alignment is such that all the photons remain inside the system for the longest time. The key difference is that the reflected photons from the bottom mirror now travel toward the top mirror instead of the beam splitter. This configuration ensures the light remains trapped within the system for more back-and-forth reflections.
\clearpage
\section{Conclusion}
\noindent While we have focused on exploiting polarization as the degree of freedom, which is not a narrow selection for future work. The no-return photon measurement method discussed here is fundamentally immune to external disruptions and other light source noise. It is scalable (for instance, multiple systems operating in parallel or in the frequency domain for flash LiDAR-like applications), and has the potential for on-chip integration for a compact form factor \cite{Steiner2021}. The proposed measurement scheme can be explored for applications in other scientific fields, including imaging, displacement measurement, bio-imaging, etc. The time-energy entangled photons \cite{Tittel2000}, and time-bin encoding could be useful in estimating the instant of a hit in second photon correlation measurements when the first photon is absorbed in the target. These investigations and explorations are reserved for our future endeavors.
\\

\noindent We do not establish any claim of ranging measurements without the return of photons, the measurement schemes discussed here assumes the biased collapse of wavefunction which needs further theoretical investigation and experimental verification. The goal of these discussions presented here is important in motivating the scientific community to explore whether such a phenomenon of information exchange or interaction is possible. We believe that the intriguing problem of no return photon ranging and imaging, along with our attempt to address it and establishing the motivation as presented here, will inspire readers to invent new ideas and ingenious measurement schemes for other scientific applications.

\section{Funding}
We want to thank the Robert A. Welch Foundation (grant A-1261), the DARPA PhENOM program, the Air Force Office of Scientific Research (Award No. FA9550-20-10366), and the National Science Foundation (Grant No. PHY-2013771). This material is also based upon work supported by the U.S. Department of Energy, Office of Science, Office of Biological and Environmental Research under Award Number DE-SC-0023103, DE-AC36-08GO28308. 

% \section{Acknowledgment}
% We thank Prof. Girish Aggarwal, Dr. Anatoly and Prof. Zubairy for their insightful discussions. We also express our gratitude to Prof. Marlan O. Scully for the valuable discussions during this research.

% \section{Additional information}
% Supplement material is included as a separate file for additional information. 

\section{Disclosures}
The authors declare that they have no known competing financial interests or personal relationships that could have appeared to influence the work reported in this paper.

\section{Data Availability}
Data sharing is not applicable to this article as no new data were created or analyzed in this study.
% Data underlying the results presented in this paper are not publicly available at this time but can be obtained from the corresponding author upon reasonable request.

%%%%%%%%%%%%%%%%%%%%%%% References %%%%%%%%%%%%%%%%%%%%%%%%%

%%%%%%%%%% using BibTeX:
\bibliography{sample}

\end{document}